# Unusual magnetic and charge transport properties in In-Substituted Half-Metallic Kagome Ferromagnet $Co_3Sn_2S_2$

Karan Singh, Subhadip Pradhan, K. Mukherjee, Ashis Kumar Nandy, Subhendra D. Mahanti, D. Topwal*

**Abstract**

The kagome ferromagnet $Co_3Sn_2S_2$ has been studied extensively for its unusual topology of electronic bands, origin of ferromagnetism and strong coupling between magnetism and charge transport. To understand the role of nonmagnetic element Sn, we have investigated magnetic, transport, and electronic structure of the isostructural compound $Co_3SnInS_2$, where all the Sn (divalent) atoms in the $Co_3Sn$ Kagome layer are replaced by In (trivalent) atoms. We find long-range ferromagnetic order is nearly quenched in $Co_3SnInS_2$. The system exhibits predominantly antiferromagnetic correlations with only a very small net magnetic moment and turns ferromagnetic in the presence of an external magnetic field. Transport measurements show a semiconducting behaviour at low temperatures. Magnetoresistance shows a nonmonotonic field dependence, changing from negative to positive with increasing magnetic field. An anomalous Hall effect is observed, but its magnitude is significantly reduced compared to $Co_3Sn_2S_2$ where the topological character of the Fermi surface plays a dominant role. These results indicate that substitution of Sn by In in the $Co_3Sn$ plane not only suppresses the topological electronic features of the transport electrons but drives the system away from the ferromagnetic Half-metallicity to an almost nonmagnetic semiconducting state with unusual paramagnetic response. Electronic structure calculations are consistent with some of these observations.

# 1. Introduction

The interplay of magnetism, electronic structure, and charge/energy transport has attracted extensive interest in the condensed matter physics community with widespread implications in both fundamental physics and technology [1-3]. More recently, the topology of the electronic band structure and resulting anomalous transport properties have become an exciting research area. Shandite $Co_3Sn_2S_2$ is an example; it exhibits long-range ferromagnetic order below $T$c ~ 178 K and large anomalous Hall effect (AHE) in the time reversal (TR) symmetry broken phase. Large AHE originates from the unusual topology of its electronic band structure near the Fermi energy [4-6]. Broken TR symmetry and spin-orbit coupling (SOC) are the key ingredients for developing large AHE [5, 6]. Several studies have been reported on this compound using spectroscopic, magnetic, and electrical transport techniques [7-11].

The origin of the ferromagnetic ground state in this system has been a subject of debate. Magnetic Co atoms form 2-dimensional $Co_3Sn$ Kagome layers (referred to as Co planes). Within a localized picture of Co moments, it was proposed that the ferromagnetic (FM) ground state arose from interplanar FM exchange and antiferromagnetic (AFM) nearest neighbor in-plane exchange couplings [12]. In a subsequent paper it was proposed that in addition to these exchange couplings, there is also a significant "across-hexagon" in-plane FM exchange which is responsible for stabilizing the long-range FM order [13]. The in-plane AFM exchange is weak but leads to frustration in the Kagome lattice and can indeed give complex magnetic structures in this compound [7, 13].

As discussed above, magnetism in this compound originates from the Co-$d$ orbitals and it has been suggested that it resides in the Co atom and that Sn and S atoms do not carry much magnetism [14]. Co has $d^7$ electrons and a simple argument suggests that in a localized picture, there should be 5 electrons in the spin-up state and 2 electrons in the spin-down state, giving rise to a large magnetic moment (1.5$\mu_B$)/Co atom. However, experimental observations reveal a much smaller net magnetic moment of 0.3$\mu_B$/Co, indicating the need for a deeper understanding of the nature of magnetism [15,16]. Guguchia et.al have suggested that the simultaneous presence of both AFM and FM ordering (a ferrimagnet) is the origin of observed reduced magnetic moment [15].

In contrast to the purely localized spin model, itinerant theories give a different physical picture of the nature of magnetism in this compound. Based on the density functional theory for the ground state and different type of response theories, Soloviev et. al argue that the magnetism of $Co_3Sn_2S_2$ has a dual nature and bears certain aspects of both itineracy and localization [17].

On the one hand, the magnetism is soft, where the local magnetic moment in the ground state is ~0.33 $\mu_B$/Co, and strongly depends on temperature and the angles formed by these moments at different Co sites of the Kagome lattice, as expected for itinerant magnets. On the other hand, the picture of localized spins still remains adequate for the description of the local stability of the ferromagnetic (FM) order with respect to the transversal spin fluctuations. The theoretical calculations predict this system to be a three-dimensional ferromagnet with the strongest exchange interaction between next-nearest neighbor Co atoms in the adjacent Kagome planes. Furthermore, the ligand states are found to play a very important role by additionally stabilizing the FM order.

Finally, in contrast to the two above homogeneous magnetic models, it has also been suggested that the decrease in FM domain size (inhomogeneous model) with cooling is the main reason for this compound's observed small net magnetic moment [16].

In addition to the magnetic and transport properties of the pristine compound, several studies have been done where chemical substitutions are made at different sites to understand the effect of these substitutions on these properties [18, 19]. In ref [18], it is reported that Ni substitution at the Co-site reduces $T_c$, but the AHE remains intact because of the persistence of the topological character of the band structure. Guguchia et al. reported that the $T_c$ is decreased with indium (In) substitution at the Sn-site and $T_c \sim 0$ for $Co_3SnInS_2$ [19]. In addition, a series of complex magnetic phases (FM, AFM, Helimagnetic) and two quantum critical points (at $T$ = 0 K) at In concentrations 0.15 and 0.5 have been seen [19]. For $Co_3SnInS_2$, the ground state was found to be paramagnetic but detailed magnetic and electrical transport measurements in this compound were not reported. Furthermore, the changes in the electronic structure (deviation from the rigid band picture) and the topology of the Fermi surface were not investigated in depth.

Clearly, a careful study of $Co_3SnInS_2$ and a detailed comparison with both $Co_3Sn_2S_2$ and $Co_3In_2S_2$ can help to unravel the nature of magnetic correlation and interplay of magnetism, transport and topological characteristics of the band structures in these three compounds because they both have the same crystal structure (but different lattice parameters), and different valence electron counts associated with the nonmagnetic elements Sn and In. In this paper we focus primarily on the magnetic and electronic transport properties of $Co_3SnInS_2$ and its electronic band structure using *ab initio* density functional theory (DFT) calculations. In addition, in the Supplementary material we give a comparative discussion of all the three compounds.

We give a brief summary of our main results first and discuss the details in following sections. In agreement with earlier studies, long-range ferromagnetic order is suppressed in $Co_3SnInS_2$. However, its magnetic properties appear to be unusual. Instead of a simple paramagnetic response, we observe antiferromagnetic correlations below ~400K. Zero field cooled (ZFC) data shows an antiferromagnetic transition neat 20 K. There is a large difference between ZFC and Field cooled data and the system develops ferromagnetic order by a small external magnetic field (~1T). Electrical resistivity measurements at low temperatures reveal semiconducting behaviour along with a change from negative to positive magnetoresistance with increasing magnetic field. The overall magnitude of the magnetoresistance however is small, below 1% compared to $10^3$ % in $Co_3Sn_2S_2$. In addition, we observe an anomalous Hall effect with a very small anomalous Hall angle of approximately 0.025%. Our DFT calculations confirm the absence of topological features near the Fermi energy in $Co_3SnInS_2$. Instead, the ground state of the system is a nonmagnetic semiconductor, in agreement with our experimental observation.

## 2. Experimental procedure

Polycrystalline $Co_3SnInS_2$ sample was synthesized using a solid-state reaction: the well-grounded stoichiometric mixtures of Co lumps, Sn powder, and S flakes were palletized and sintered inside evacuated quartz tubes at 750 ◦C for 72 h. The phase purity of the sample was investigated using the Rigaku Smart Lab instrument with Cu$K\alpha$ source. Rietveld refinement of the XRD data was performed using FullProf Suite software. Temperature (*T*) and magnetic field (*H*) dependent magnetization were performed using a Magnetic Property Measurement System (MPMS). Electrical Transport and Hall Effect measurements were carried out using the Physical Property Measurement System (PPMS); both are from Quantum design. In addition to $Co_3SnInS_2$, for comparative study we also synthesized $Co_3Sn_2S_2$ and $Co_3In_2S_2$ and measured their physical properties.

## 3. Computational details

Density functional theory (DFT) was used to obtain the electronic structure of these compounds, using the VASP code [20]. All calculations were performed using the primitive rhombohedral unit cell of $Co_3SnInS_2$, where In occupies the Wyckoff 1a site (0, 0, 0), Sn occupies the 1b site (1/2, 1/2, 1/2), Co occupies the 3d sites (1/2, 0, 0), (0, 1/2, 0), and (0, 0, 1/2), and S occupies the 2c sites.. The lattice parameters and internal atomic positions were fully relaxed until the total-energy change was below $1 \times 10^{-5}$ eV/atom and the residual forces on each atom were less than $5 \times 10^{-3}$ eV/°A. The exchange–correlation energy was treated within the generalized gradient approximation (GGA) using the Perdew–Burke–Ernzerhof (PBE) functional [21]. The interaction between valence electrons and ionic cores was described using the projector augmented-wave (PAW) method [22], with PAW datasets corresponding to Co, Sn d, In d, and S. A plane-wave kinetic-energy cutoff of 500 eV was used to ensure convergence of total energies and forces. Brillouin-zone integrations were performed using a Γ-cantered $12 \times 12 \times 12$ **k**-point mesh. Spin-polarized calculations were performed to explore the magnetic nature of the ground state assuming that the spins are collinear. In these calculations we did not include Spin-Orbit coupling (SOC). However, in our calculations including SOC noncollinear formalism was allowed.

## 4. Results and discussion

### A. Experimental results

Fig. 1a shows the Rietveld refined XRD pattern for $Co_3SnInS_2$ at room temperature. Like $Co_3Sn_2S_2$, this compound crystallizes in a hexagonal unit cell with space group *R-3m* and is in a single phase. The crystal structure and Kagome's lattice view are drawn from the Rietveld refined XRD pattern using Vesta software (fig.1b and 1c). The obtained crystallographic parameters are given in Supplementary Table 1. Compared to $Co_3Sn_2S_2$, the lattice parameter of $Co_3SnInS_2$ increases along the *c*-direction and decreases along the *a*- and *b*-directions due to the difference in ionic radii of Sn and In.

Fig. 2a shows the temperature-dependent zero-field-cooled (ZFC) and field-cooled (FC) magnetization (*M*) of for H=0.01T for temperatures in the range 4-390K. *M* decreases with temperature and below 380K, ZFC–FC data start to differ and a small slope change occurs near $T_1 \sim 300$ K (inset fig. 2a) in both. For T < ~50 K, *M* starts to increase with decreasing T and ZFC data clearly shows an AF type transition around T =$T_2 \sim 23$ K, although $T_2 \sim 0$K in the case of FC measurement. As shown in Fig. 2(b), with the increasing magnetic field (0.5–6 T) the low-temperature transition $T_2$ is suppressed, while $T_1$ shifts to higher temperatures (inset of Fig. 2(b)). We find that replacing half the Sn atoms by In in $Co_3Sn_2S_2$ suppresses the ferromagnetic transition at $T_c$ (Supplementary Fig. S2a) and leads to two new anomalies at temperatures $T_1$ and $T_2$. Our measurements are in agreement with the results of ref [19] who found a decrease of $T_c$ with In-substitution in $Co_3Sn_{2-x}In_xS_2$ and $T_c = 0K$ in $Co_3SnInS_2$.

In $Co_3SnInS_2$, we observed a magnetic moment of ~ 0.006 $\mu_B$/f.u at 2K under ZFC, which is much smaller than the 0.3 $\mu_B$/f.u. observed in $Co_3Sn_2S_2$ (see fig. 2a and supplementary figs. S2a and S2b). This result combined with the temperature dependence of the magnetization below 23K (ZFC measurement) suggests that the ground state of $Co_3SnInS_2$ has a complex magnetic nature.

We also measured the temperature-dependent magnetization $M$ of the end compound ($Co_3In_2S_2$) under an applied field of $H = 0.01T$ over the temperature range 4–390 K (see Supplementary Fig. S2b). The magnetization decreases with decreasing temperature, and below 380 K ZFC and FC curves start to split. It is noted that the magnetic moment reaches ~ 0.0106 $\mu_B$/f.u. at 2 K under ZFC $M$, which is larger than that of $Co_3SnInS_2$ but smaller than that of $Co_3Sn_2S_2$.

Next, we measured the field dependence of $M$ at 2 K and 300 K (fig. 2c). At 2 K, $M$ increases with field, shows hysteresis at low fields (inset fig. 2c), and saturates with magnetic moment 0.17 $\mu_B$/f.u. around 1.5 T, with a reduced moment compared to $Co_3Sn_2S_2$ for which the saturation moment is 0.76 $\mu_B$/f.u. (Supplementary Fig. S2c). At 300 K, $Co_3Sn_2S_2$ is paramagnetic (see supplementary Fig. S2c), while $Co_3SnInS_2$ exhibits field-induced saturation of magnetic moment 0.14 $\mu_B$/f.u. at ~ 0.85 T. We also studied the $Co_3In_2S_2$, whose isothermal magnetization behavior is similar to that of $Co_3SnInS_2$, but its saturated moments (~ 0.45 $\mu_B$/f.u. at 2K) is larger than $Co_3SnInS_2$ (see supplementary fig. S2c).

Fig. 2d gives the electrical resistivity ($\rho_{xx}$) of $Co_3SnInS_2$ which shows clear anomalies at $T_1$ and $T_2$. Upon cooling from 390 K, $\rho_{xx}$ increases and reaches a maximum near $T_1$, then decreases with further cooling. At $T_2 \sim 23$ K, $\rho_{xx}$ exhibits a minimum before increasing again down to 2 K. These anomalies are consistent with the magnetization results measured at 0.01 T where we also saw changes in magnetic response. Even at a magnetic field of 9 T, the overall temperature dependence of $\rho_{xx}$ remains unchanged, although its magnitude decreases by a small amount. Both $T_1$ and $T_2$ appear in $\rho_{xx}$ even at 9 T. Since magnetization changes very little at $T_2$ but the temperature dependence of resistivity changes dramatically near $T_2$, it is most likely that $T_2$ anomaly is electronic (non-magnetic) in origin. The resistivity at 2 K in zero magnetic field is 5.2 mΩ·cm, which is two orders of magnitude larger than 0.01 mΩ·cm observed in $Co_3Sn_2S_2$ and $Co_3In_2S_2$ (see supplementary fig. S3). Thus, electronic transport in this series changes from metallic to semiconducting at the 50% Indium substitution and then back to metallic behaviour with 100% Indium substitution.

The magnetoresistance, $MR = \frac{\rho(T,B) - \rho(T,0)}{\rho(T,0)}$, of Co$_3$SnInS$_2$ between 2 and 50 K is shown in Fig. 2e, where $\rho(T,0)$ and $\rho(T,B)$ are the resistivities at zero and finite magnetic field $B$. At 2 K, $MR$ decreases with field, reaching a minimum at $B_{min}$ ~ 3 T, and then increases, reaching ~0.96% at higher fields. The magnitude $MR$ in Co$_3$SnInS$_2$ is much smaller than in Co$_3$Sn$_2$S$_2$ (where $MR$ ~10$^3$ %, see supplement fig. S4a, is associated with the presence of topological Weyl nodes near the Fermi level). Extremely smaller $MR$ in Co$_3$SnInS$_2$ suggests that 50% In-substitution moves the Weyl nodes away from the Fermi level.

With increasing temperature, $B_{min}$ shifts to higher fields and the magnitude of $MR$ also decreases. We fitted the $MR$ with equation [23]:

$$MR = \alpha B + \gamma B^2 \quad \text{.......} \quad (1)$$

where $\alpha$ and $\gamma$ are the coefficients shown in Fig. 2f. The coefficient of the linear term (α) is negative, while coefficient of the quadratic term (γ) is positive. This suggests that the low-field negative $MR$ arises from the suppression of electron-spin fluctuation scattering by the external magnetic field. In contrast, the positive $MR$ varying quadratically with B can arise from different sources such as a two-carrier model or distribution of mobilities of carriers with energies distributed about the Fermi energy (deviations from the Kohler rule, [24])

We also measured the $MR$ in Co$_3$Sn$_2$S$_2$ and Co$_3$In$_2$S$_2$, which show similar behaviour. In Co$_3$Sn$_2$S$_2$, $MR$ is positive follows a $B^{1.5}$ dependence (see supplement fig. S4b). In Co$_3$Sn$_2$S$_2$, negative $MR$ is observed near $T_c$ because near this temperature scattering from spin fluctuations dominates and supressing this scattering by the external field reduces the resistivity. In Co$_3$In$_2$S$_2$, $MR$ exhibits a $B^2$ dependence at 2 K (see supplement fig. S6), consistent with the behaviour seen in simple metallic system.

Figure 3a presents the Hall resistivity ($\rho_{xy}$) of Co$_3$SnInS$_2$ measured at various temperatures (2-100 K). $\rho_{xy}$ can be expressed as the sum of ordinary ($\rho_{xy}^O$) and anomalous ($\rho_{xy}^A$) contributions [25]:

$$\rho_{xy} = \rho_{xy}^O + \rho_{xy}^A \quad \text{........} \quad (2)$$

where $\rho_{xy}^O = R_0 B$ and $\rho_{xy}^A = R_s M$ ($R_0$ and $R_s$ are the constant of proportionality). In the high-field regime ($B > 1$ T), where $M$ saturates, we fit Eq. (2) to extract the ordinary Hall coefficient ($R_0$, slope) and anomalous Hall contribution ($R_s M$, intercept). From $R_0$, the ordinary Hall term ($\rho_{xy}^O$) and anomalous Hall term ($\rho_{xy}^A = \rho_{xy} - \rho_{xy}^O$) were obtained (Fig. 3b). In this compound, the carrier concentration is hole-doped ($n_h = 1/(R_0 e)$) and is of the order of ~$10^{21}$ cm$^{-3}$, showing weak temperature dependence (Fig. 3c). This carrier concentration is similar to that of Co$_3$Sn$_2$S$_2$ (~$10^{21}$ cm$^{-3}$, hole-doped) but about one order of magnitude smaller than in Co$_3$In$_2$S$_2$ (~$10^{22}$ cm$^{-3}$, also hole-doped).

Fig. 3d shows the anomalous Hall resistivity ($\rho_{xy}^A$). At 2 K, $\rho_{xy}^A$ increases sharply with field, saturates near 1.4 µΩ·cm, and exhibits hysteresis at low fields (Fig. 3e), consistent with magnetization (see inset Fig.2c). With increasing temperature, both the spontaneous and saturated values of $\rho_{xy}^A$ decrease (Fig. 3e). The anomalous Hall resistivity $\rho_{xy}^A$ in Co$_3$SnInS$_2$ is much smaller than in Co$_3$Sn$_2$S$_2$ (~16 µΩ·cm at 2K, see supplementary Fig. S5). In addition, no anomalous Hall contribution is observed in Co$_3$In$_2$S$_2$ (see inset of supplementary Fig. S6).

The anomalous Hall effect arises from the combined effects of magnetization and spin–orbit coupling and contains both intrinsic, extrinsic and impurity scattering contributions [25]. The intrinsic part originates from the Berry curvature of the electronic bands, while the extrinsic part is due to skew scattering and side-jump scattering. However, in doped sample there are also chemical disorder in sample which introduce the impurity scattering [26]. The relative importance of these mechanisms, and the size of anomalous Hall effect, depends on types of scattering, and the electronic structure near the Fermi level. In-substitution shifts the Weyl nodes away from the Fermi level, suppressing the topological contribution to the anomalous Hall effect; therefore, either extrinsic or impurity scattering mechanism dominates the anomalous Hall response in Co$_3$SnInS$_2$.

To identify the underlying origin of $\rho_{xy}^A$, we calculated the anomalous Hall conductivity ($\sigma_{xy}^A$):

$$\sigma_{xy}^A = \rho_{xy}^A / (\rho_{xy}^2 + \rho_{xx}^2) \quad \ldots\ldots\ldots\ldots (3)$$

Figure 4a shows $\sigma_{xy}^A$ as a function of magnetic field at different temperatures. At 2 K, $\sigma_{xy}^A$ reaches a maximum value of 0.03 Ω$^{-1}$·cm$^{-1}$ and remains finite at zero field with a value of 0.029 Ω$^{-1}$·cm$^{-1}$. The temperature dependence of $\sigma_{xy}^A$ and $\sigma_{xx} (= \rho_{xx}/(\rho_{xy}^2 + \rho_{xx}^2))$ at 0 T and 9

T is shown in Figs. 4b and 4c, respectively. At 2 K, the anomalous Hall conductivity ($\sigma_{xy}^A$) and longitudinal conductivity ($\sigma_{xx}$) are of the order of 0.03 $\Omega^{-1}\cdot cm^{-1}$ and 190 $\Omega^{-1}\cdot cm^{-1}$ at 0 T, and 0.05 $\Omega^{-1}\cdot cm^{-1}$ and 188 $\Omega^{-1}\cdot cm^{-1}$ at 9 T, respectively. These values are comparable to those reported for EuTiO$_3$ ($\sigma_{xy}^A = 0.1$ $\Omega^{-1}\cdot cm^{-1}$; $\sigma_{xx} = 250 \Omega^{-1}\cdot cm^{-1}$ [26]), suggesting that $\sigma_{xy}^A$ is primarily induced due to impurity scattering in the sample. Next, we calculated the anomalous Hall angle ($\sigma_{AH}$):

$$\sigma_{AH} = \frac{\sigma_{xy}^A}{\sigma_{xx}} \quad \ldots\ldots\ldots.. (4)$$

Figure 4d shows the temperature dependence of the anomalous Hall angle ($\theta_{AH}$), calculated using Eq. (3). At 2 K, $\theta_{AH}$ is 0.015% at 0 T and 0.025% at 9 T, which is significantly smaller than the ~20% reported for Co$_3$Sn$_2$S$_2$ [6]. This strongly indicates that In-substitution removes the topological contribution associated with Weyl nodes in Co$_3$Sn$_2$S$_2$. The magnitudes of $\sigma_{xy}^A$ and $\sigma_{xx}$ in Co$_3$SnInS$_2$ suggest that the anomalous Hall effect mainly arises from impurity scattering combined with weak spin-orbit coupling. [26].

**B. Theoretical Results**

Our electronic-structure calculations indicate that the band structure in the neighborhood of the Fermi energy and the nature of the ground state changes dramatically when 50% of the Sn sites in Co$_3$Sn$_2$S$_2$ are replaced by In. Co$_3$SnInS$_2$ becomes a narrow-gap semiconductor and the topological points move far away from the Fermi energy. These changes can be understood from two things. One there are significant changes in the band structure because the Co atoms in the Kagome plane have surrounding In atoms instead of Sn and rigid band picture is not valid. Secondly, the electron counting: indium has one fewer valence p electron than tin (In: [Kr] 4d$^{10}$5s$^2$5p$^1$; Sn: [Kr] 4d$^{10}$5s$^2$5p$^2$), reducing the total electron count by one per substituted atom. As a result, the linear band crossings (Weyl-type crossings in the absence of SOC) that lie close to the Fermi level in Co$_3$Sn$_2$S$_2$ (see Supplementary Fig. S8(b)) move away from the Fermi energy. If this was all and a rigid band picture was valid then Co$_3$SnInS$_2$ would have been a metal. Instead, it becomes a narrow band gap (gap ~0.23 eV) semiconductor with the valence band maxima (VBM) at the T and L points of the BZ and the conduction band minima (CBM) at the $\Gamma$ and T points. These results are consistent with previous reports [27, 28]. SOC plays an important role, in its absence the VBM is at the L point and the CBM is at the $\Gamma$ and T points.

Collinear spin-polarized calculations without including SOC (see Fig. 1(b)) show that the Co magnetic moment, which is about 0.3 $\mu_B$/Co atom in $Co_3Sn_2S_2$, is strongly suppressed and becomes essentially zero in $Co_3SnInS_2$. This trend is consistent with earlier theoretical reports [27]. Experimentally, a small residual magnetization has been reported in this compound; however, such a weak moment may not be captured within GGA and could require beyond-DFT treatments (e.g., hybrid functionals, or GW) to describe correlation effects more accurately. Within GGA, the 1 Sn→In substitution drives the system from a ferromagnetic metal to a nonmagnetic narrow-gap semiconductor.

The small anomalous Hall conductivity experimentally observed at low temperature in $Co_3SnInS_2$ is more likely attributed to external defects which introduce holes into the system and extrinsic mechanisms (such as skew scattering or side-jump contributions) rather than an intrinsic Berry-curvature-driven response of the band structure. Figure 1(c) shows the band structure along the high symmetry k-path (See Supplementary Fig. S7(b)), computed self-consistently including spin–orbit coupling (SOC) within the noncollinear formalism. The dashed horizontal line denotes the Fermi level at 0 eV. The bands responsible for the linear Weyl crossings and the associated strong Berry curvature in $Co_3Sn_2S_2$ are shifted to energies above ~1 eV in $Co_3SnInS_2$. The Co-$d_{x2-y2}$ band, which lies well below $E_F$ in $Co_3Sn_2S_2$, shifts close to the Fermi level in $Co_3SnInS_2$ and is therefore expected to have a strong impact on the transport properties of this compound. The density of states (DOS), shown in Fig. 1(a), confirms that Co 3d states dominate near the Fermi energy. An lm-decomposed DOS for the Co 3d states (inset) shows that the electronic states near $E_F$ are dominated by Co 3d-$x^2-y^2$ and 3d-xz orbitals, and the important role of orbital character (and the kagome-lattice framework) in shaping the states relevant for transport.

Although the present constrained spin-polarized DFT calculations yield a nonmagnetic narrow band-gap semiconducting ground state, qualitatively consistent with our experiments, the true nature of the ground state and low-energy excitations in $Co_3SnInS_2$ may be considerably more complex. As shown in Fig. 5, the occupied Co-3d bands near the Fermi level are very narrow, with bandwidth W<0.5 eV. Such a small W implies a reduced kinetic energy scale, so that the ratio U/W (where U is the on-site Coulomb interaction) can become large. When U/W□1, electronic correlations are expected to be significant and are not adequately captured within standard DFT. The temperature dependence of the magnetization between 380 K and 100 K (Fig. 2a) shows strong antiferromagnetic correlations, further supporting the importance of electron–electron interactions. These observations suggest that $Co_3SnInS_2$ may lie in a strongly correlated regime, possibly exhibiting pseudogap behavior. A proper description of its low-energy physics would therefore require approaches beyond conventional DFT, such as Hubbard-type effective Hamiltonians treatments.

## 5. Conclusions

In conclusion, our comprehensive magnetic and electrical transport investigations on the isostructural compound $Co_3SnInS_2$ demonstrate that In-substitution strongly modifies the magnetic and electronic ground state of the parent Weyl ferromagnet $Co_3Sn_2S_2$. The long-range ferromagnetic order is suppressed and shows complex antiferromagnetic correlation with a small residual magnetic moment, accompanied by semiconducting transport behavior. Our measurements indicate that In-substitution suppresses the intrinsic Weyl-related transport mechanism characteristic of $Co_3Sn_2S_2$, such as very large MR. The evolution from negative magnetoresistance at low magnetic fields to positive magnetoresistance at higher fields (small in magnitude), together with the presence of an anomalous Hall effect, indicates the competing roles of different scattering processes (spin-dependent and spin-independnet). Furthermore, the anomalous Hall angle is significantly reduced compared to the parent compound, pointing to enhanced impurity scattering and weak spin–orbit coupling. Consistent with our experimental findings, theoretical calculations give a nonmagnetic semiconducting ground state. The Weyl nodes in the neighbourhood of the Fermi energy, which are responsible for the large magnetoresistance move far away from the Fermi level upon 50% Indium substitution. Overall, this study highlights the sensitivity of the magnetic correlations and electronic structure to chemical substitution and establishes $Co_3SnInS_2$ as an important platform to further understand the evolution of magnetism and topological effects in the $Co_3Sn_{2-x}In_xS_2$ family.

**AUTHOR INFORMATION**

**Corresponding author**

**[D. Topwal: dinesh.topwal@iopb.res.in]**

**Authors**

**Karan Singh** - *Institute of Physics, Sachivalaya Marg, Bhubaneswar 751005, India*

**Subhadip Pradhan** - *School of Physical Sciences, National Institute of Science Education and Research, An OCC of Homi Bhabha National Institute, Jatni 752050, India*

**K. Mukherjee** - *School of Basic Sciences, Indian Institute of Technology Mandi, Mandi 175005, Himachal Pradesh, India*

**Ashis Kumar Nandy** - *School of Physical Sciences, National Institute of Science Education and Research, An OCC of Homi Bhabha National Institute, Jatni 752050, India*

**Subhendra D. Mahanti** - *Department of Physics and Astronomy, Michigan State University, East Lansing, MI 48824, USA*

**D. Topwal** - *Institute of Physics, Sachivalaya Marg, Bhubaneswar 751005, India*

*and*

*Homi Bhabha National Institute, Training School Complex, Anushakti Nagar, Mumbai 400094, India*

**Notes**

The authors declare no competing financial interest.

**Data availability**

The data supporting this study's findings are available from the corresponding author upon reasonable request

**Acknowledgements**

S.P. and A.K.N. acknowledge the computational resources, Kalinga cluster, at the National Institute of Science Education and Research, Bhubaneswar, India. K.M acknowledges IIT Mandi for experimental facilities.

**Figures**

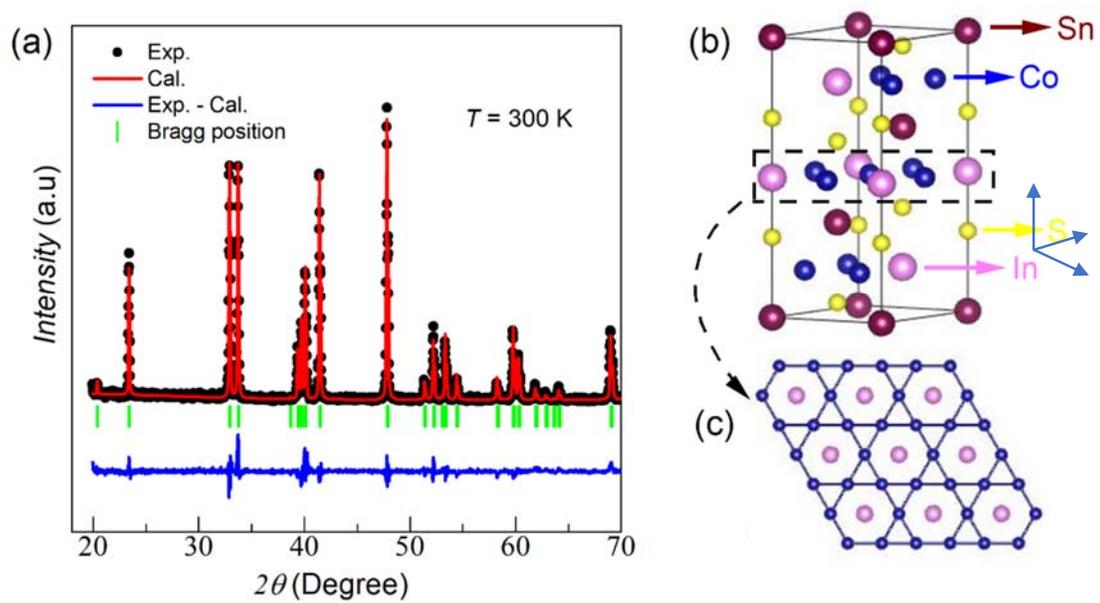

**Fig.1:** (a) show the Rietveld refined X-ray diffraction (XRD) pattern at 300 K. (b) show the crystal structure (hexagonal unit cell) for the $Co_3SnInS_2$. (c) show the two-dimensional kagome lattice.

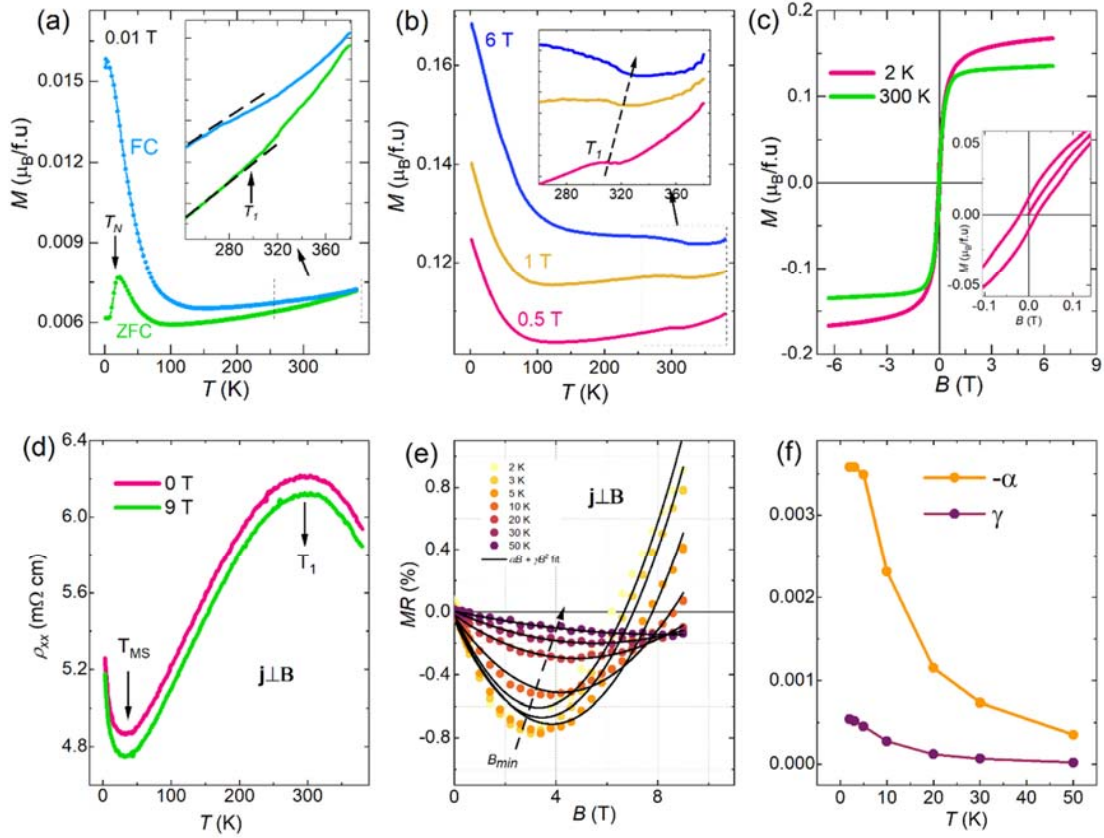

**Fig.2:** (a) Temperature (*T*) dependent magnetization (*M*) measured at 0.01 T under ZFC (zero-field cooled) and FC (field cooled). Inset: Zoomed region in temperature range 240-380 K displaying transition $T_1$. (b) *T* dependent *M* measured at different field under ZFC. Inset: Zoomed *M* in high temperature region showing shift in $T_1$ to higher temperature with increasing magnetic field. (c) Isotherm *M* at 2 and 300 K measured in a complete loop of magnetic field. Inset: low-field region showing magnetic hystresis at 2 K. (d) *T* dependent longitudinal resistivity $(\rho_{xx})$ measured at 0 and 9 T. (e) Magnetoresistance (*MR*) plotted at different temperatures. The black solid lines represent fits to the equation (1). (f) Fitting coefficients as a function of temperature.

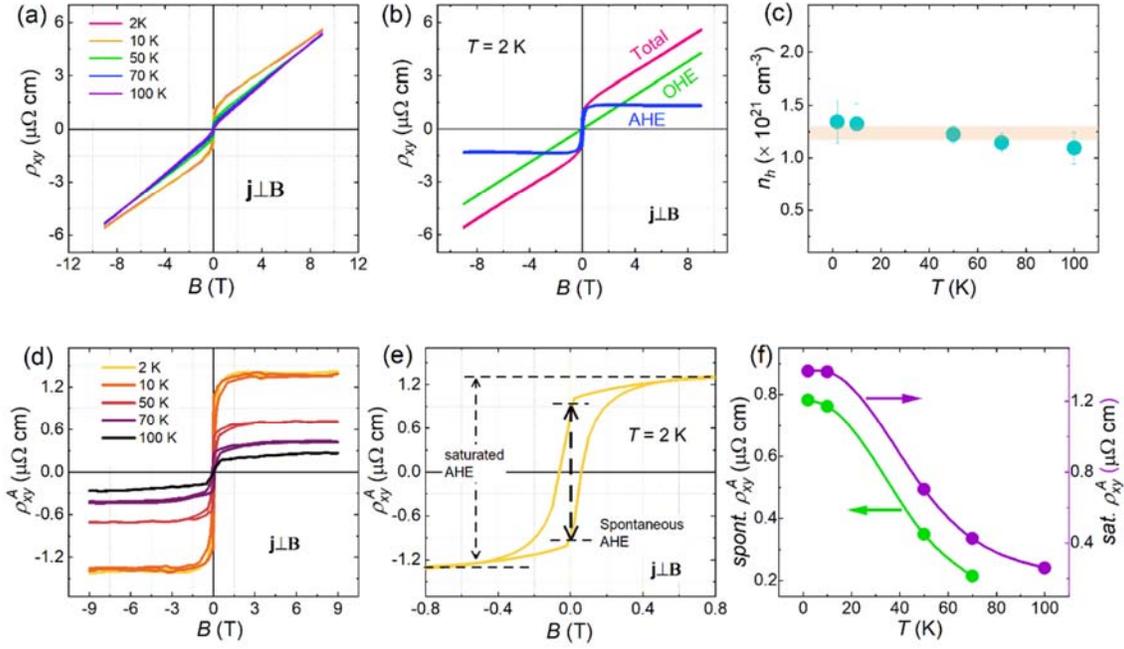

**Fig.3:** (a) Magnetic field dependent Hall resistivity ($\rho_{xy}$) measured at different temperatures. (b) Magnetic field dependent ordinary Hall effect (OHE) and anomalous Hall effect (AHE) extracted from the total $\rho_{xy}$. (c) Temperature dependent carrier concentration ($n_h$). (d) Magnetic field dependent anomalous Hall resisitivity ($\rho_{xy}^A$) plotted at different temperatures. (e) $\rho_{xy}^A$ at 2 K showing the saturated and spontaneous AHE. (f) Temperature dependent spontaneous $\rho_{xy}^A$ (left-axis) and saturated $\rho_{xy}^A$ (right-axis).

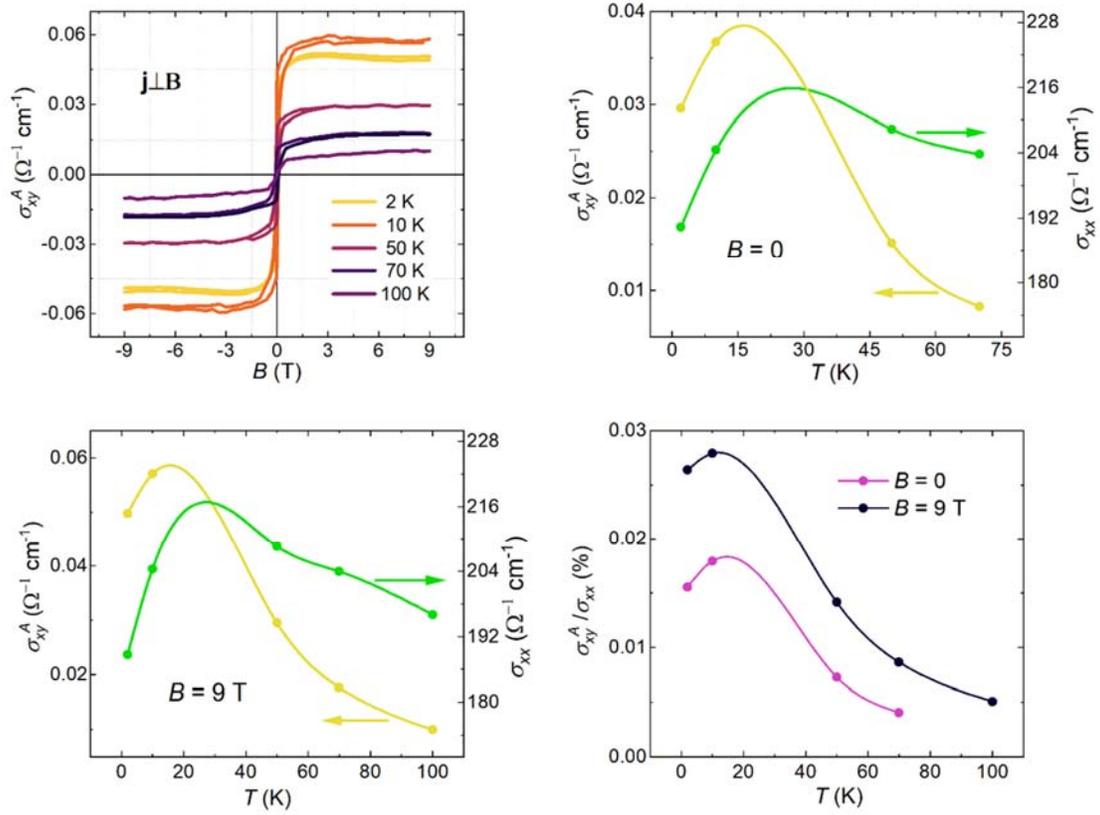

**Fig.4:** (a) Magnetic field dependent (a) anomalous Hall conductivity ($\sigma_{xy}^A$). Temperature dependent $\sigma_{xy}^A$ (left-axis) and longitudinal conductivity $\sigma_{xx}$ (right-axis) at (b) $B = 0$ and (c) $B = 9$ T. (d) Temperature dependent Hall ratio ($\sigma_{xy}^A/\sigma_{xx}$) for $B = 0$ and 9 T.

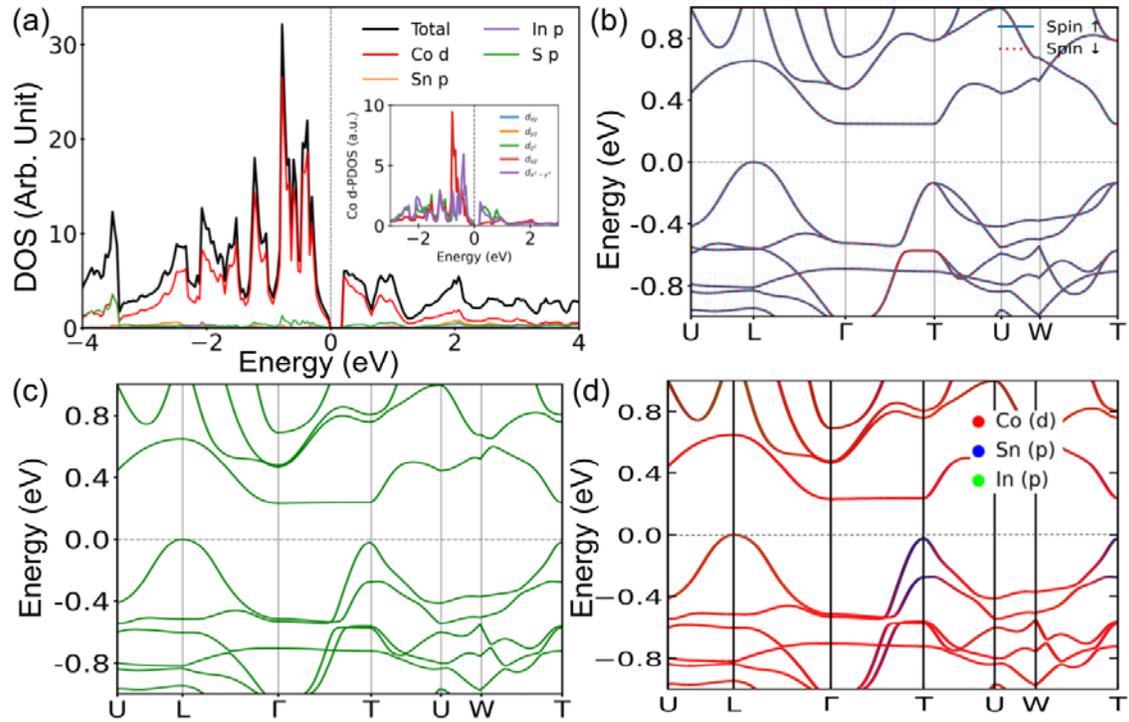

**Fig.5:** Electronic structure of $Co_3SnInS_2$ from GGA calculation. (a) Total and orbital projected partial density of states (DOS); The inset shows the different Co 3d orbitals contribution near EF. (b) Spin-polarized band structure (spin-up and spin-down) along the high-symmetry path in 3D BZ, showing nearly identical dispersions for the two spin channels (negligible exchange splitting). (c) Band structure including spin–orbit coupling (SOC). (d) Orbital-resolved band structure with SOC, highlighting the dominant contributions from Co 3d states (red) and the smaller Sn/In p-state weights (blue/green).

# Supplementary materials

## for

## Anomalous Electrical Transport in In-Substituted Half-Metallic Kagome Ferromagnet $Co_3Sn_2S_2$

Karan Singh[1], Subhadip Pradhan[2], K. Mukherjee[3], Ashis Kumar Nandy[2], Subhendra D. Mahanti[4], D. Topwal[1,5]

[1]*Institute of Physics, Sachivalaya Marg, Bhubaneswar 751005, India*

[2]*School of Physical Sciences, National Institute of Science Education and Research, An OCC of Homi Bhabha National Institute, Jatni 752050, India*

[3]*School of Physical Sciences, Indian Institute of Technology Mandi, Mandi 175075, Himachal Pradesh, India*

[4]*Department of Physics and Astronomy, Michigan State University, East Lansing, MI 48824, USA*

[5]*Homi Bhabha National Institute, Training School Complex, Anushakti Nagar, Mumbai 400094, India*

**Crystal structure**

Figures S1a and S2b show the Rietveld refined XRD pattern for the $Co_3Sn_2S_2$ and $Co_3In_2S_2$ at room temperature, respectively. Both samples crystallize in a rhombohedral structure (space group, $R$ -$3m$)

**Magnetization**

Figure S2a shows the temperature dependent zero-field cooled (ZFC) and field cooled (FC) magnetization ($M$) of the $Co_3Sn_2S_2$ at magnetic field of 0.01 T. With decreasing temperature, $M$ increases and starts to show bifurcation between ZFC and FC below $T_c \sim$ 179 K; this temperature corresponds to the long-range ordered ferromagnetic transition temperature, and the bifurcation could cause by out-of-plane ferromagnetic exchange interaction which agrees well with the earlier reports [1-3]. In cooling down further we observe a second transition at 130 K in the real ($\chi'$) and imaginary ($\chi''$) parts of the AC susceptibility at 831 Hz and 0.01 T (inset of figure S2a). A similar transition has been reported for the same compound by other authors [4]. The competition between the in-plane AF and out-of-plane ferromagnetic exchange interactions are responsible for the two magnetic phase transitions (at 130 K and 179 K) [3].

Figure S2b depicts the temperature-dependent ZFC and FC $M$ measured at 0.01 T for $Co_3In_2S_2$. It is noticed that $M$ curves bifurcate about 390 K and the it decreases with lowering temperatures. The AC susceptibility result indicates no features (not shown) across the whole range of recorded temperatures. These findings show that both transitions (at 179 and 130 K) are suppressed, while antiferromagnetic correlation persist below 390K.

Next, we measured the isotherm $M$ at 2K for $Co_3Sn_2S_2$ and $Co_3In_2S_2$, as shown in Figure S2c. For $Co_3Sn_2S_2$, $M$ increases rapidly and reaches saturation at 0.7 T. In $Co_3In_2S_2$, $M$ increases with $H$, although not as abruptly as in $Co_3Sn_2S_2$, and reaches saturation at 1.5 T, which is a higher field than in $Co_3Sn_2S_2$. In the low-field region, weak hysteresis is observed in both compounds (upper inset fig. S2c). The saturation moment in $Co_3In_2S_2$ is less than that of $Co_3Sn_2S_2$. At 300 K (lower inset fig. S2c), in $Co_3Sn_2S_2$, $M$ increases linearly with $H$ without saturation, showing paramagnetic behaviour. In $Co_3In_2S_2$, $M$ grows with $H$ and reaches saturation at same field (1.5 T) as found in 2K.

**Electrical transport**

Figure S3a show the temperature-dependent resistivity ($\rho_{xx}$) at 0 T for $Co_3Sn_2S_2$ and $Co_3In_2S_2$. For $Co_3Sn_2S_2$, in the zero field, the room temperature $\rho_{xx}$ is 1.5 mΩ cm which decreases to 0.0187 mΩ-cm by 2 K, giving a residual resistivity ratio (RRR) of ~ 80. In addition, a kink is also observed at the $T_c$. Above inset of figure S3a shows the $T^{2.4}$ dependent $\rho_{xx}$ at low temperature (T<~150K), which changes to a linear $T$ dependence at higher temperatures. The low-$T$ behaviour suggests a significant contribution of the electron–magnon scattering in addition to the electron-electron scattering [5]. For $Co_3In_2S_2$, it is observed that $\rho_{xx}$ decreases with decreasing temperature and reaches to 0.638 mΩ cm and 0.00543 mΩ cm at 300 K and 2 K respectively. The RRR (~ 117) is much larger than the $Co_3Sn_2S_2$ sample, indicating a strong metallic character. Below inset of figure S2a shows the linear dependent of $\rho$ with $T^{1.8}$, indicating the dominance of electron-electron scattering ($\rho \sim T^2$). The electron-magnon scattering is not important. As $Co_3Sn_2S_2$ has been extensively investigated by magnetoresistance (*MR*) and Hall effect [2], which show a giant anomalous Hall effect due to the presence of large Berry curvature. Our *MR* and Hall effect measurements also support the anomalous electronic properties. Details of our measurements are presented in the Fig. S4 and Fig. S5.

In figure S6, we show the field dependence of *MR* at 2K for $Co_3In_2S_2$. *MR* increases with *H* and reaches a value of 2 at 9 T, indicating conventional metallic character. For further study, we have measured the Hall resistivity ($\rho_{xy}$) at 2 K (inset of figure S6). It is seen that $\rho_{xy}$ increases linear with $\rho_{xy} > 0$ for $H > 0$ and $\rho_{xy} < 0$ for $H < 0$, indicating ordinary Hall effect (OHE). It arises due to the deflection of charge carriers by the Lorentz force. It is given by the equation: $\rho_{xy} \sim R_0 B$, where $R_0$ is the ordinary Hall coefficient whose sign depends on the nature of the charge carriers (electron-like or hole-like). The positive $\rho_{xy}$ with positive direction of magnetic field (and vice-versa) indicates the dominance of the hole like carriers of the order of $1.5 10^{22}$ $cm^{-3}$ in this system which is 10 order larger than $Co_3Sn_2S_2$ and $Co_3SnInS_2$ ($n_h$ $10^{21}$ $cm^{-3}$). From this study, we conclude that in $Co_3In_2S_2$, the electronic properties are associated with the electronic bands and do not depend on the observed magnetization. This may be because the occupied bands contributing to the magnetization are far away from the Fermi level. As a result, these electrons do not strongly influence the charge carriers around the Fermi level. Large AHE associated with the Berry curvature of the bands is also suppressed in $Co_3In_2S_2$.

**Electronic structure**

**Co₃Sn₂S₂**

Figure S8a shows the band structure of $Co_3Sn_2S_2$ in the paramagnetic state without SOC. As we will discuss in detail, there are three bands close to the Fermi energy (chosen as 0 eV) in the energy +/- 0.5 eV, which we will denote as bands 1, 2, and 3 respectively. These three bands control the magnetic and transport properties of $Co_3Sn_2S_2$. Band 1 (width ~0.375 eV along U - L - Γ) is flat along Γ - T indicating nearly 2-d character (confined to the *xy* plane), and band 2 (width ~0.5 eV) is the one which is just above band 1 at the Γ point dispersing downwards as one goes from Γ to L and then disperses upwards as one goes from Γ to T. Band 2 has a local maximum at the L point. Band 3 is highly dispersive along Γ – T, but has a dispersion of ~0.4 eV along T-U-W-T. Bands 1 and 2 have two crossings at ~ 0.125 eV and 0.245 eV above the Fermi energy. These are potential Weyl nodes and are not important for transport in the paramagnetic state. In the ferromagnetic phase (Fig. S8b), bands 1, 2, and 3 are spin split, the blue bands are for the spin-up and the red bands are for the spin-down states. The exchange splittings are ~0.2-0.5 eV. Spin up bands are occupied and the spin down bands are empty, indicating a half-metallic ground state. The linear band crossing (Weyl nodes) for the spin up bands are at -0.01 eV, and 0.085 eV, respectively, close to the Fermi energy, and these Weyl nodes (resulting in large Berry curvature) can dramatically affect the charge transport properties leading to large anomalous Hall effect [2]. When SOC is included, small energy gaps (~0.04 eV) open up at the Weyl nodes with band anti crossings (Fig. S8c). These features are in full agreement with the results of previous studies on this compound [2]. The orbital projected bands are shown in Fig. S8d, indicating that bands 1 and 3 are mostly Co-*d* and band 2 is a hybridized band between Co-*d* and Sn-*p*, but again mostly Co-*d*. If one looks at the density of states (DOS) (fig. S7c) near the Fermi energy $E_F$, there are two narrow bands of widths ~0.5 eV, and $E_F$ lies in the middle of the lower energy band and the system is metallic. In the FM state, this lower energy band is filled with spin-up electrons, and the system becomes a half-metal. The calculated Co atomic magnetic moment for $Co_3Sn_2S_2$ is 0.34μB/Co, in excellent agreement with earlier calculations [8] and experiment [2].

**Co₃In₂S₂**

When all the Sn atoms are replaced by In atoms, two things happen at an atomic level. The Sn *s*, *p* valence orbitals (atomic energies -13.04 eV and -6.76 eV) are replaced by In *s*,*p* valence orbitals (energies –10.14 eV and -5.37 eV) indicating that bonding with Co-*d* (energy ~ -17.77 eV) gets reduced. Secondly, in fact more importantly, the valence count reduces from 4 to 3.

If we assume that the band structure does not change in going from Sn to In, then the narrow band near the Fermi energy gets emptied and the In compound should be a paramagnetic semiconductor. But there are significant changes in the band structure (compare Fig. S8a and Fig. S91a) and $Co_3In_2S_2$ becomes a metal with rapidly changing (decreasing) density of states near the Fermi energy, which makes it an interesting system in its own right, as seen in our experimental measurements. In fact, the hole-like carriers seen in Hall measurements can be easily understood using our calculated band structure.

In the work presented in Ref. 8, the authors investigated the physical properties of $Co_3In_{2-x}Sn_xS_2$ (x=0,2) compounds, focusing on their electronic, magnetic, and thermoelectric behaviors. However, a detailed comparison of the band structure near the chemical potential was not made in Ref. 8. The major changes that occur in going from the Sn to In system are: (i) Band 1 gets emptied from U-L-Γ-T and the part from T-U-W-T which was completely filled (below the Fermi energy) is partially filled, (ii) Band 2 is far above and does not play any role in magnetism and transport, (iii) Different parts of the Band 3 which is 2-fold degenerate contributes to the states near $E_F$. As a result, $Co_3In_2S_2$ is not a topological semimetal but is most likely a strongly interacting metal where exchange interactions can play an important role in magnetism. However, spin polarized (collinear) DFT calculations give the ground state of this compound as a weak ferromagnet with magnetic moment 0.11μB/Co, nearly 1/3 of $Co_3Sn_2S_2$. So, changing the environment of Co from Sn to In decreases the strength of the ferromagnetic magnetic moment. The calculated moment agrees with the results reported in the ref [8]. Although constrained DFT gives a ferromagnetic state with very small moment for the In system, the true ground state of this system appears to be much more complex. The increase in magnetization with $T$ starting from 2K until 380K (Fig. S2b) suggests some sort of non collinear spin structure with antiferromagnetic correlations that decrease with increasing temperature. To summarize our theoretical results, there are three Co-$d$ bands that control the magnetic and transport properties of $Co_3(Sn, In)_2S_2$ system. In the Sn compound, part of Band 1 and part of Band 3 which are within 0.5 eV of $E_F$ are responsible for the observed strong ferromagnetism. Furthermore, in the ferromagnetic state, the spin-up Band 1 intersects Band 2 giving linear band crossings and Weyl physics [2]. In the In compound, magnetism is determined by different parts of Band 1 and Band 3 which were considerably below $E_F$ in the Sn compound but become active in the In compound.

**Table 1: Unit cell parameter and atomic positional parameter for $Co_3Sn_2S_2$ and $Co_3SnInS_2$**

| Lattice parameters of $Co_3Sn_2S_2$ | | | |
|---|---|---|---|
| $a$ (Å) | $b$ (Å) | $c$ (Å) | $V$ (Å$^3$) |
| 5.36(7) | 5.36(7) | 13.17(5) | 378.36 |
| Lattice parameters of $Co_3SnInS_2$ | | | |
| $a$ (Å) | $b$ (Å) | $c$ (Å) | $V$ (Å$^3$) |
| 5.32(2) | 5.32(2) | 13.67(6) | 386.89 |
| Lattice parameters of $Co_3In_2S_2$ | | | |
| $a$ (Å) | $b$ (Å) | $c$ (Å) | $V$ (Å$^3$) |
| 5.31(4) | 5.31(4) | 13.65(8) | 384.87 |

| Atomic positions | | | | |
|---|---|---|---|---|
| | site | $x$ | $y$ | $z$ |
| Co | 9e | 0.5 | 0 | 0 |
| Sn1(or Sn or In1) | 3a | 0 | 0 | 0 |
| Sn2(or In or In2) | 3b | 0 | 0 | 0.5 |
| S | 6c | 0 | 0 | z |

**Supplementary Figures**

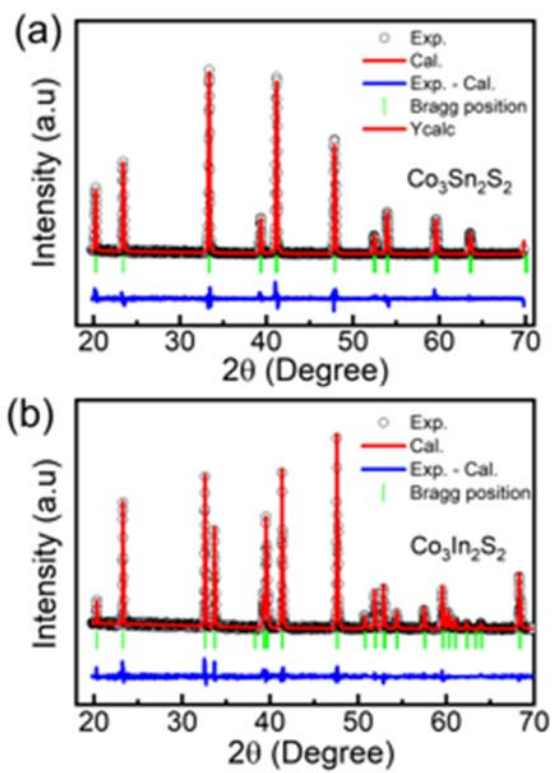

**Fig. S1**: (a) and (b) show the Rietveld refined X-ray diffraction (XRD) pattern at 300 K for $Co_3Sn_2S_2$ and $Co_3In_2S_2$, respectively.

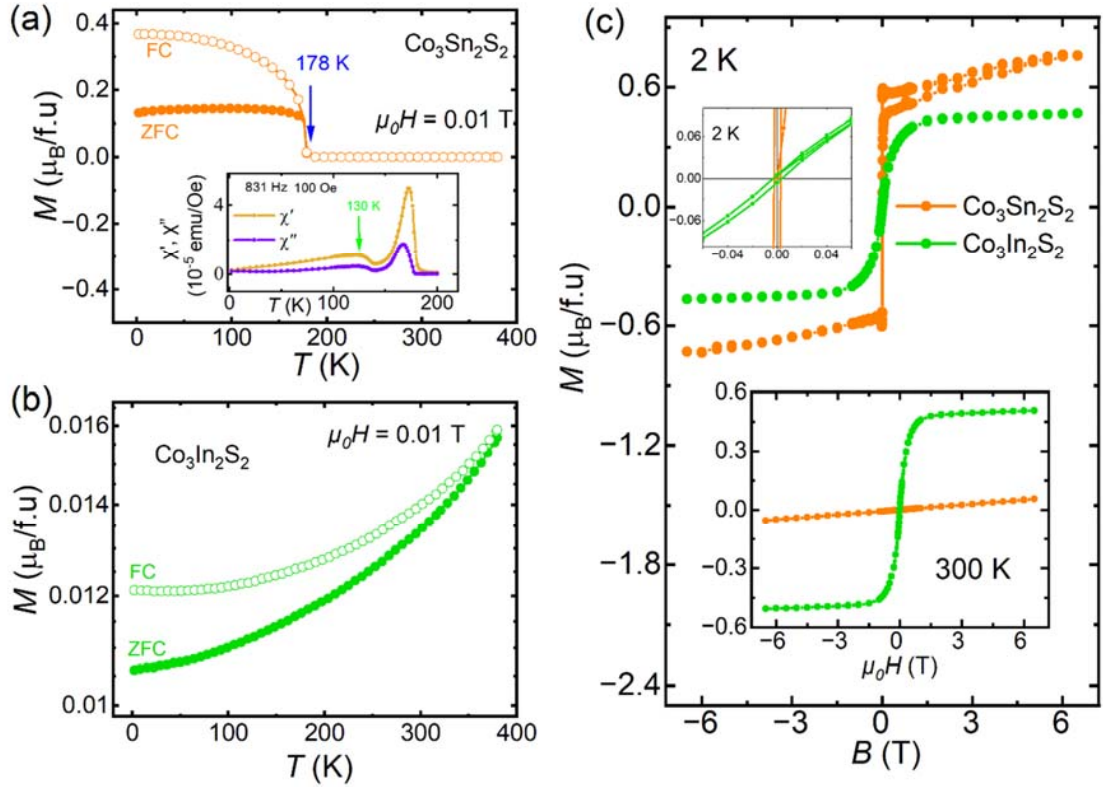

**Fig. S2:** (a) Temperature dependent magnetization (M), zero field cooled (ZFC) and field cooled (FC) at 0.01 T for $Co_3Sn_2S_2$. Lower inset shows the temperature dependent AC susceptibility at 0.01 T and 831Hz. (b) Temperature dependent magnetization ($M$), zero field cooled (ZFC) and field cooled (FC) at 0.01 T for $Co_3In_2S_2$. (c) Magnetic field dependent magnetization at 2K for $Co_3Sn_2S_2$ and $Co_3In_2S_2$. Upper inset: Zoomed same fig. at low fields. Lower inset: magnetic field dependent magnetization at 300K for $Co_3Sn_2S_2$ and $Co_3In_2S_2$

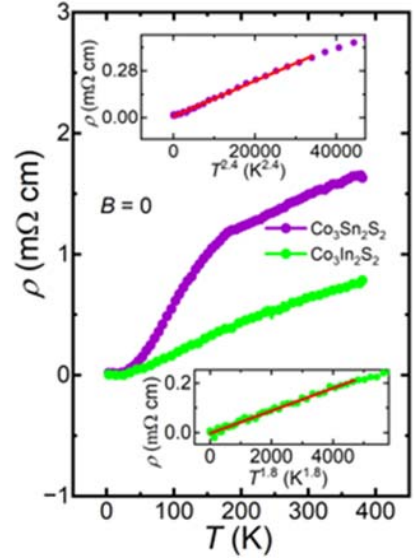

**Fig. S3**: (a) Temperature dependent resistivity ($\rho$) at 0 T for $Co_3Sn_2S_2$ and $Co_3In_2S_2$. Upper inset: $T^{2.4}$ dependent $\rho$ for $Co_3Sn_2S_2$. Lower inset: Inset figure: $T^{1.8}$ dependent $\rho$ for $Co_3In_2S_2$. Red curve shows the linear fitting for both compounds.

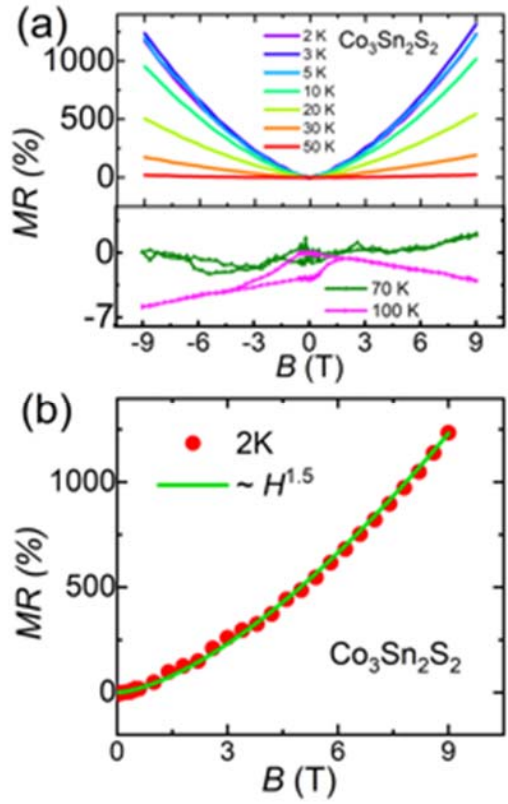

**Fig. S4**: (a) Magnetic field dependent magnetoresistance (*MR*) at different temperatures $Co_3Sn_2S_2$. (b) *MR* plotted at the 2K. Green line is the fit to the equation $MR \sim H^{1.5}$. The value of MR increases and reaches a value of about 1280% at 2 K and 9 T.

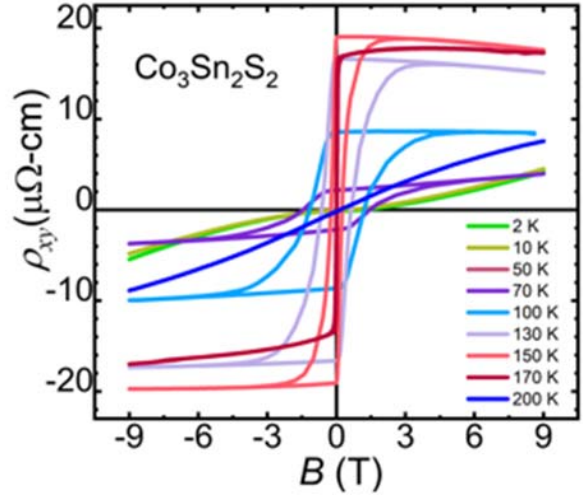

**Fig. S5:** Magnetic field dependent Hall resistivity ($\rho_{xy}$) at different temperatures for $Co_3Sn_2S_2$.

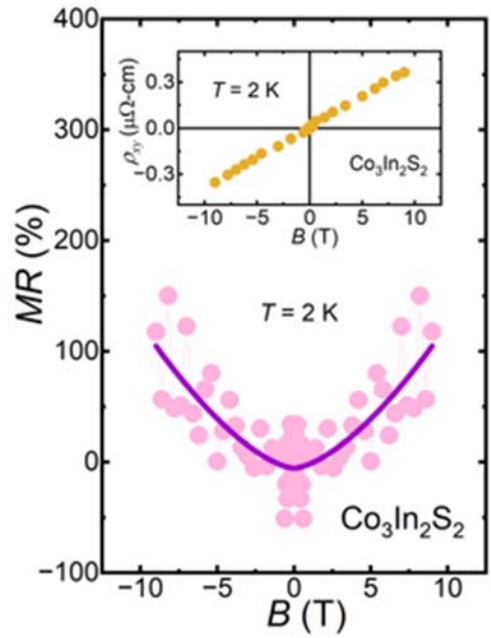

**Fig. S6:** Magnetic field dependent magnetoresistance (*MR*) at 2K for $Co_3In_2S_2$. Inset: Magnetic field dependent Hall resistivity ($\rho_{xy}$) at 2 K for $Co_3In_2S_2$.

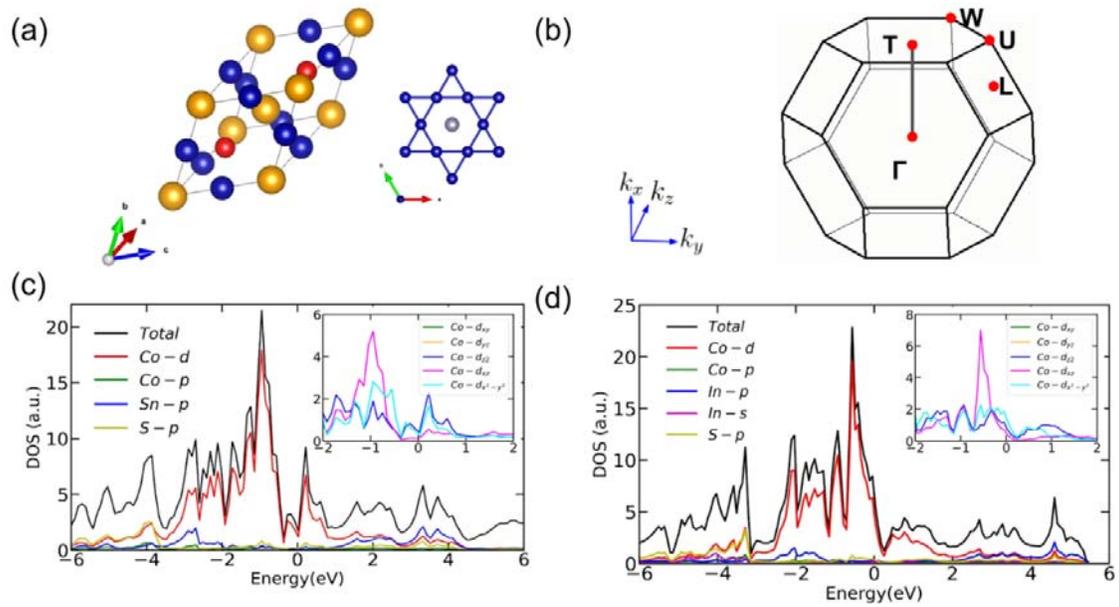

**Fig. S7:** : (a) primitive unit cell (Navy = Co, Yellow = Sn/In and Red = S). subfigure displays the Kagome' lattice structure of $Co_3Sn$ layer. (b) 3-D Brillouin zone (BZ). The calculated partial density of states (PDOS) of (c) $Co_3Sn_2S_2$ (b) $Co_3In_2S_2$ including SOC. Inset: shows Co-*3d* orbital contributions.

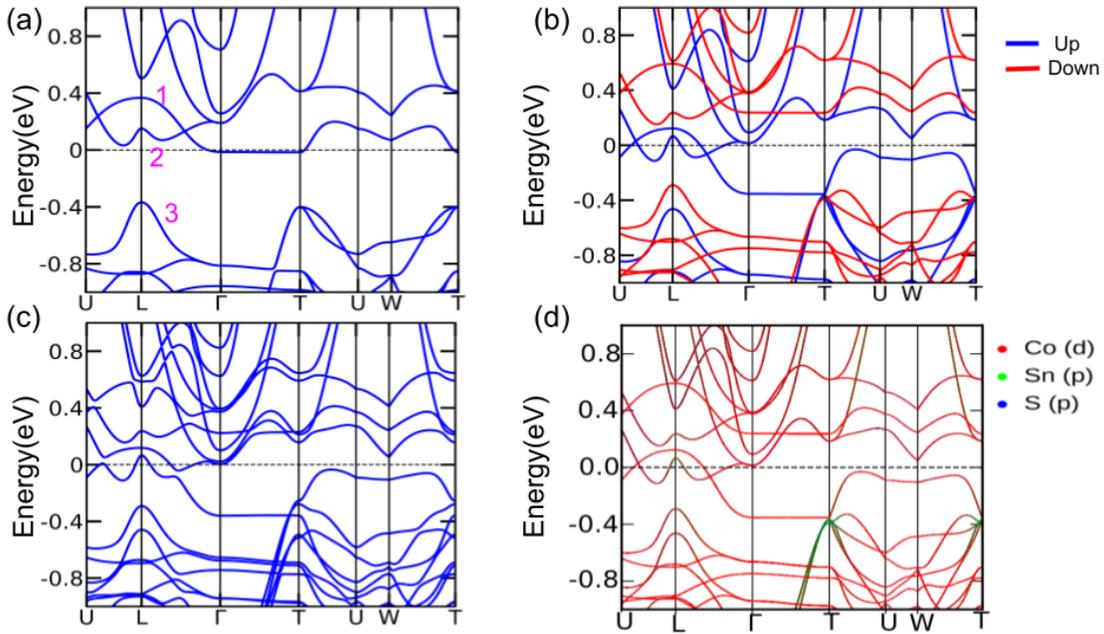

**Fig. S8:** Band structure of Co$_3$Sn$_2$S$_2$ along the high symmetry K-path for (a) Paramagnetic state (b) Magnetic state without Spin-orbit coupling (WSOC), The orange color is for the spin up and the dotted blue curve for spin down (c) For SOC case. (d) The orbital projected band plots in WSOC bands.

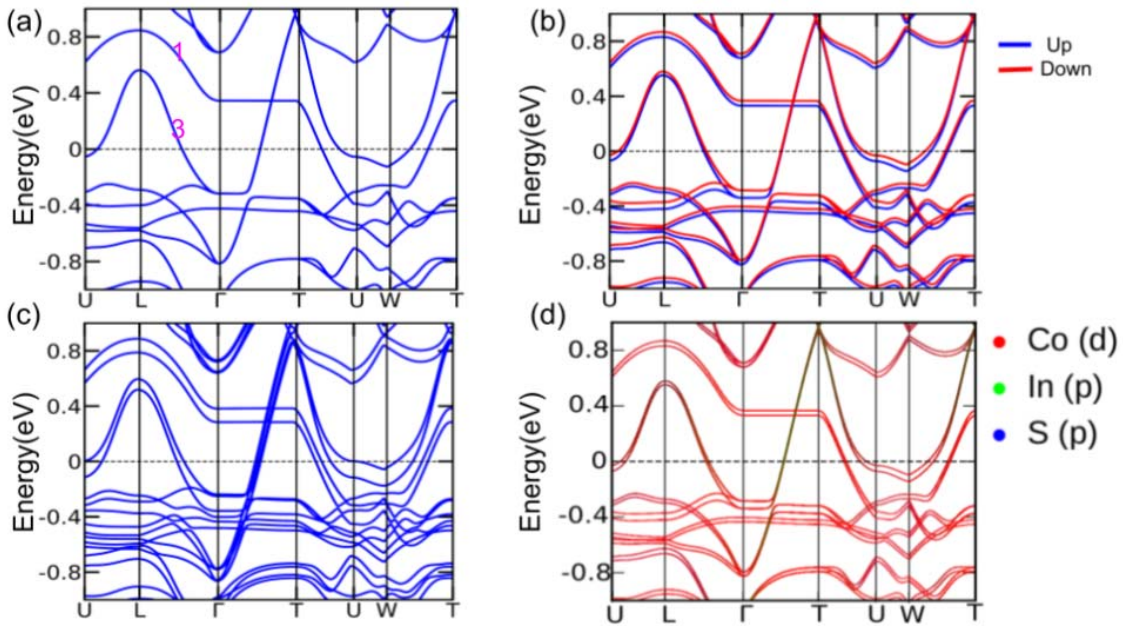

**Fig. S9:** Band structure of Co$_3$In$_2$S$_2$ along the high symmetry K-path for (a) Paramagnetic state (b) Magnetic State without Spin-orbit coupling (WSOC), The orange color is for the spin up

and the dotted blue curve for spin down (c) For SOC case. (d) The orbital projected band plots in WSOC bands.